\documentclass[twocolumn,amsmath,amssymb,prd]{revtex4}
\usepackage{amsfonts}
\usepackage{mathrsfs}
\usepackage{bbm}
\usepackage{amsmath}
\usepackage{amssymb}
\usepackage{dcolumn}
\usepackage{hyperref}
\usepackage{graphicx}
\usepackage{epstopdf}
\usepackage{latexsym}
\usepackage{subfigure}
\usepackage{bm}               % bold math
\usepackage{color}
\usepackage{ulem}

\begin{document}
\setlength{\parskip}{0pt}
\centerline{} \vskip 10mm

\title{Reconstruction of Gravitational Lensing Using WMAP 7-Year Data}

\author{Chang Feng}
\email{cfeng@physics.ucsd.edu}
\affiliation{Center for Astrophysics
and Space Sciences, University of California San Diego, La Jolla, CA
92093}

\author{Brian Keating and Hans P. Paar}
\affiliation{Center for Astrophysics and Space Sciences and the Ax
Center for Experimental Cosmology, University of California San
Diego, La Jolla, CA 92093}
\author{Oliver Zahn}
\affiliation{Berkeley Center for Cosmological Physics and
Lawrence Berkeley Laboratory, University of California, Berkeley,
CA 94720}

\begin{abstract}
Gravitational lensing by large scale structure introduces
non-Gaussianity into the Cosmic Microwave Background and imprints a
new observable, which can be used as a cosmological probe. We apply
a four-point estimator to the Wilkinson Microwave Anisotropy Probe
(WMAP) 7-year coadded temperature maps alone to reconstruct the
gravitational lensing signal. The Gaussian bias is simulated and
subtracted, and the higher order bias is investigated. We measure a
gravitational lensing signal with a statistical amplitude of
$\mathcal {C}$ = $1.27\pm 0.98$ using all the correlations of the W-
and V-band Differencing Assemblies (DAs). We therefore conclude that
WMAP 7-year data alone, can not detect lensing.
\end{abstract}

\pacs{98.80.Cq; 98.80.-k}

\maketitle

\section{introduction}

Gravitational lensing of the Cosmic Microwave Background (CMB)
provides information on the mass distribution between the surface of
last scattering and the observer, thus potentially providing
information, for example, on dark energy and neutrino masses. In
addition, gravitational lensing causes $E$-modes to be converted
into large angular scale $B$-modes, thereby potentially
contaminating $B$-mode signature of inflationary gravitational waves
\cite{kks}. Because lensing deflects CMB photons by approximately
$3'$, a perturbative treatment to first order is generally valid. An
estimator for the deflection angle has been devised by Hu
\cite{waynehuTri,waynehuTT}.

The first attempt to detect lensing by Hirata et al.
\cite{Hirata2004} used the cross-correlation between the WMAP 1-year
data and selected luminous red galaxies (LRGs) from the Sloan
Digital Sky Survey (SDSS). No statistically significant signal was
found. The first detection of lensing was performed by Smith et al.
\cite{smith2007} who used the cross-correlation between the NRAO VLA
Sky Survey (NVSS) of radio galaxies with a higher mean redshift than
the Sloan LRGs and a fully-optimal lensing estimator on the
statistically more powerful WMAP 3-year data. Evidence for lensing
was found at the 3.4$\sigma$ level. Using a similar estimator as in
\cite{Hirata2004}, Hirata et al. \cite{hirata2008} obtained results
consistent with, though at slightly lower significance than
\cite{smith2007}, using WMAP 3-year data, LRGs and quasars from the
SDSS data, as well as data from the NVSS. Recently, Smidt et al.
\cite{smidt2010} used an estimator based upon the kurtosis of the
CMB temperature four-point correlation function to estimate lensing
from WMAP 7-year data only and claimed evidence for lensing at the
$2 \sigma$ level. Recently, the Atacama Cosmology Telescope (ACT)
collaboration successfully detected gravitational lensing
\cite{ACTlensing} at the 4$\sigma$ level. The South Pole Telescope
(SPT) detected the effects of gravitational lensing on the angular
power spectrum\cite{SPT}.

In this paper we present a search for gravitational lensing using
the WMAP 7-year data alone and the standard optimal quadratic
estimator \cite{waynehuTri,waynehuTT} which differs from the
kurtosis estimator of \cite{smidt2010}. We apply the quadratic
estimator to WMAP-7 temperature maps alone for the first time in the
hopes that our analysis might serve as a touchstone allowing for
consistent comparison between different lensing extraction
techniques. We review the notation for full-sky reconstruction of
gravitational lensing in Section II. We discuss the sky-cut used in
our analysis in Section III. Then we introduce our modified
estimator in Section IV making use of the optimal quadratic
estimator of \cite{waynehuTri}. We introduce the WMAP 7-year data in
Section V, and describe the details of the calculations, including
the noise model, and analysis in Section VI. Results of a null test
are shown in Section VII, and we discuss the conclusions of our work
in Section VIII.

\section{Gravitational Lensing}

The effect of lensing on the CMB's primordial temperature
$\tilde{T}$ in direction ${\bf n}$ can be represented by
\begin{equation}
T(\mathbf{n})=\tilde{T}(\mathbf{n}+\mathbf{d}(\mathbf{n})),
\label{cmblensing}
\end{equation}
where $T$ is the lensed temperature and $\mathbf{d}(\mathbf{n}) =
\mathbf{\nabla} \phi$, with $\phi$ being the lensing potential.
% We use the terms two-point correlation and four-point correlation and reserve the term powerspectrum
% for the factor C that multiplies the delta-function.
The two-point correlation function of the temperature field
following \cite{hufull}, is:
\begin{eqnarray}
\langle T_{lm} T_{l'm'} \rangle = \tilde{C}^{TT}_l\delta_{ll'}\delta_{m-m'} (-1)^m + \hspace{15mm} \nonumber \\
\displaystyle\sum_{LM}(-1)^M \left(
\begin{array}{ccc}
l   &  l'  & L\\
m & m'&-M
\end{array} \right) f^{TT}_{lLl'} \phi_{LM}, \label{lensingcouplemode}
\end{eqnarray}
where the second term encodes the effects of lensing with the
weighting factor $f^{TT}_{lLl'}$ given by
\begin{eqnarray}
f^{TT}_{lLl'} = \tilde{C}_l^{TT} {} _0F_{l'Ll} + \tilde{C}_{l'}^{TT}
{} _0F_{lLl'}. \label{weighting}
\end{eqnarray}
Here $\tilde{C}_l^{TT}$ are the unlensed temperature power spectra.
and
\begin{eqnarray}
_0F_{lLl'} = \sqrt{\frac{(2l+1)(2l'+1)(2L+1)}{4\pi}} \times \hspace{15mm} \nonumber\\
\frac{1}{2} [L(L+1)+l'(l'+1)-l(l+1)] \left(
\begin{array}{ccc}
l  &  L &l'\\
0 & 0 & 0
\end{array} \right).
\end{eqnarray}

The lensing estimator is constructed from an average over a pair of
two-point correlations \cite{waynehuTri,waynehuTT} and has the form
\begin{eqnarray}
d^{TT}_{LM} = \frac{A^{TT}_L} {\sqrt{L(L+1)}} \times \hspace{35mm} \nonumber\\
\displaystyle\sum_{ll'mm'}(-1)^Mg^{TT}_{l'l}(L)\left(
\begin{array}{ccc}
l'   &  l  &  L \\
m' & m &-M
\end{array} \right)T_{l'm'}T_{lm}.\label{TT}
\end{eqnarray}

The requirement that the estimator in Eq. (\ref{TT}) is unbiased and
has minimal variance results in
\begin{eqnarray}
A^{TT}_L = L(L+1)(2L+1) \biggl[ \displaystyle\sum g^{TT}_{ll'}(L) f^{TT}_{lLl'} \biggr]^{-1} \label{estTTnoise}
\end{eqnarray}
and
\begin{eqnarray}
g^{TT}_{ll'}(L)=\frac{f^{TT}_{lLl'}}{2C^{tot}_lC^{tot}_{l'}},
\label{TTfilter}
\end{eqnarray}
with $C_l^{\rm tot} = C_l^{TT} + N_l^{TT}$, where $C_l^{TT}$ are the
lensed power spectra and $N_l^{TT}$ is the instrumental noise. In
the following, the summations are from $l$ and $l'=0$ to 750 and
$|m|\leq l$, $|m'|\leq l'$. The WMAP 7-year data do not contain
additional information at higher multipoles.

To reduce computation time we follow \cite{hufull} and define three
maps for the TT estimator:
\begin{eqnarray}
{}_0A^T(\mathbf{n}) =
\displaystyle\sum_{lm}\frac{1}{C^{tot}_l}T_{lm} {\,}
_{0}Y_{lm}(\mathbf{n}), \label{abbrivA}
\end{eqnarray}
\begin{eqnarray}
X(\mathbf{n}) =
\displaystyle\sum_{lm}\frac{\tilde{C}^{TT}_l}{C^{tot}_l}
T_{lm}\alpha_{l0} {\,} _{+1}Y_{lm}(\mathbf{n}), \label{abbrivX}
\end{eqnarray}
\begin{eqnarray}
Y(\mathbf{n}) =
\displaystyle\sum_{lm}\frac{\tilde{C}^{TT}_l}{C^{tot}_l}T_{lm}\beta_{l0}
{\,} _{-1}Y_{lm}(\mathbf{n}), \label{abbrivY}
\end{eqnarray}
and take the inverse Spherical Harmonic Transform (SHT) of $_0A^T X$
and $_0A^T Y$ to get
\begin{equation}
\Upsilon^{(1)}_{LM} = \beta_{L0}\int d\mathbf{n} {\,}
_{+1}Y^{\ast}_{LM}{\ }_0A^T X \label{P1}
\end{equation}
\begin{equation}
\Upsilon^{(2)}_{LM} = \alpha_{L0}\int d\mathbf{n} {\,}
_{-1}Y^{\ast}_{LM}{\ }_0A^T Y \label{P2}
\end{equation}
with
\begin{equation}
\alpha_{ls} =- \sqrt{\frac{(l-s)(l+s+1)}{2}} \label{ca}
\end{equation}
\begin{equation}
\beta_{ls} = \sqrt{\frac{(l+s)(l-s+1)}{2}}. \label{cb}
\end{equation}
Using Eqs. (\ref{abbrivA}), (\ref{abbrivX}) and (\ref{abbrivY}) the
expression for $d_{LM}^{TT}$ in Eq. (\ref{TT}) becomes
\begin{equation}
d^{TT}_{LM} = \frac{A^{TT}_L} {\sqrt{L(L+1)}} \bigl[
\Upsilon^{(1)}_{LM}+\Upsilon^{(2)}_{LM} \bigr]. \label{newTT}
\end{equation}

A similar procedure is followed for the efficient calculation of
$A_L^{TT}$ in Eq. (\ref{estTTnoise}). The resulting expression is
given in \cite{smith} (originally proposed in \cite{cora}):
\begin{eqnarray}
A_l^{TT} = \int^{-1}_{+1} \bigg[ \Big[ \xi^T_{00}(\theta)
\xi^T_{11}(\theta) - \xi^T_{01}(\theta) \xi^T_{01}(\theta) \Big]
d^l_{-1-1}(\theta)
\nonumber \\
+ \Big[ \xi^T_{00}(\theta) \xi^T_{1-1}(\theta) - \xi^T_{01}(\theta)
\xi^T_{0-1}(\theta) \Big] d^l_{1-1}(\theta) \bigg] d(\cos\theta)
\hspace{2mm} \label{newnoise}
\end{eqnarray}
with the $\xi^T$ given by
\begin{equation}
\xi^T_{00}(\theta)=\displaystyle\sum_l\frac{2l+1}{4\pi}\frac{1}{C^{TT}_l+N^{TT}_l}d^l_{00}(\theta),\label{XI1}
\end{equation}
\begin{equation}
\xi^T_{0\pm1}(\theta)=\displaystyle\sum_l\frac{2l+1}{4\pi}\sqrt{l(l+1)}
\frac{\tilde{C}^{TT}_l}{C^{TT}_l+N^{TT}_l}d^l_{0\pm1}(\theta),\label{XI2}
\end{equation}
\begin{equation}
\xi^T_{1\pm1}(\theta)=\displaystyle\sum_l\frac{2l+1}{4\pi}l(l+1)
\frac{(\tilde{C}^{TT}_l)^2}{C^{TT}_l+N^{TT}_l}d^l_{1\pm1}(\theta),\label{XI3}
\end{equation}
here $d^l_{ss'}(\theta)$ are Wigner d-functions.

\section{Sky Cut}

In order to eliminate contaminated data, regions such as the
galactic plane and bright point sources in the full-sky map must be
removed using a mask, thereby introducing a sky-cut. For example, in
\cite{hirata2008}, the Kp2 mask was used to make 84.7\% of the sky
uncontaminated. In \cite{smidt2010}, the more conservative KQ75 mask
was used to clean artifacts around the galactic plane and point
sources.

The sky-cut can be removed as a separate component to get a full-sky
map before we process the data. One such technique is the
``inpainting" method in which the estimated values of pixels in the
map are substituted for those removed by the mask. Perotto et al.
have simulated the full sky reconstruction for PLANCK \cite{both}.
The full-sky map recovered in this way will bias the lensing
reconstruction slightly.

Another method proposed by A. Benoit-Levy \cite{ABL} apodizes the
masked regions of the map and inpaints the masked regions of the map
by constrained Gaussian random values of the unlensed temperature.
In this way, the sky-cut-induced coupling approximately reduces to a
unit matrix. However, for WMAP, we have to remove a big portion of
the sky, reducing $f_{\rm sky}$ dramatically to $0.3$. The unbiased
estimator could be scaled up by a factor of $1/f_{\rm sky}$, but the
signal-to-noise ratio would be reduced significantly. This means the
uncertainty of the reconstructed signal would be larger.

As opposed to a separate-component solution, we obtain an
all-inclusive lensing reconstruction pipeline, using the built-in
filter of the estimator to treat the data without pre-conditioning
it. The optimal estimator for the potential based on the maximum
likelihood is derived by Hirata \cite{hirataLikeli}. The full
inverse variance $({\bf{C}}+{\bf{N}})^{-1}$, instead of
$(C^{TT}_L+N^{TT}_L)^{-1}$, was used by \cite{smith2007} because it
is an optimal filter when there are sky-cuts and inhomogeneous
noise. The sky-cut generates artifacts in harmonic space, as does
lensing. $({\bf{C}}+{\bf{N}})^{-1}$ can be used to filter those
modes affected by the sky-cut. However, we do not use this filter
because the inversion of $({\bf{C}}+{\bf{N}})$ is computationally
challenging \cite{smith2007}; instead we use the estimator Eq.
(\ref{TT}) which is identical to the one of \cite{hirata2008}, and
it is an excellent approximation to the maximum likelihood
estimator. We note that, while $(C^{TT}_L+N^{TT}_L)^{-1}$ will be
suboptimal to a full $({\bf{C}}+{\bf{N}})^{-1}$ filter, it preserves
the simplicity and efficiency of the lensing reconstruction
procedure.
\begin{figure}
\centering
\includegraphics[width=9cm,height=9cm]{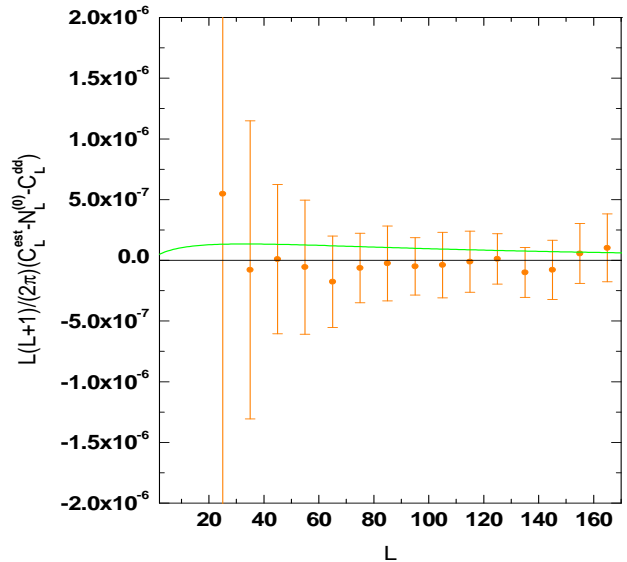}
\caption{The higher order bias calculated from $(C_{L}^{est} -
N^{(0)}_L) - C_L^{dd}$ for all correlations of the WMAP's W- and
V-band DAs. The simulated higher order bias from averaging 700 (to
be discussed in Figure \ref{converge}) realizations is shown in
orange. For comparison, the simulated lensing signal is shown in
green. } \label{higher}
\end{figure}

\section{The lensing estimator}

For WMAP, we modify the estimator slightly to deal with the
instruments' anisotropic temperature noise.

The observed lensed temperature map $\mathbb{T}$ is given by
\begin{eqnarray}
\mathbb{T}(\mathbf{n}) = M(\mathbf{n}) \big[ \int d\mathbf{n}'
T(\mathbf{n}') B(\mathbf{n},\mathbf{n}') + N(\mathbf{n}) \big] \label{beamandmask}
\end{eqnarray}
and likewise the ``observed" unlensed temperature map
$\tilde{\mathbb{T}}$ is
\begin{eqnarray}
\tilde{\mathbb{T}}(\mathbf{n}) = M(\mathbf{n}) \big[ \int d\mathbf{n}'
\tilde{T}(\mathbf{n}') B(\mathbf{n},\mathbf{n}') + N(\mathbf{n})]
\label{beamandmaskGauss}
\end{eqnarray}
Here $M(\mathbf{n})$ represents the mask, $B(\mathbf{n},
\mathbf{n'})$ the beam, and $N(\mathbf{n})$ the noise.

For a pair of maps $\alpha$ and $\beta$, ``$TT(\alpha\times\beta)$"
denotes the cross-correlation between these two temperature maps. A
harmonic mode of the reconstruction including noise is estimated as
\begin{eqnarray}
\mathbbm{d}^{TT(\alpha \times \beta)}_{LM} & = & \frac{A^{TT(\alpha
\times \beta)}_L} {\sqrt{L(L+1)}}
\displaystyle\sum_{ll'mm'}(-1)^Mf^{TT}_{lLl'} \left(
\begin{array}{ccc}
l'   &   l &  L\\
m' & m& -M
\end{array}
\right)
\nonumber \\
& \times & \frac{\mathbb{T}^{(\alpha)}_{l'm'}}
{\mathbb{C}^{(\alpha)}_{l'}} \frac{\mathbb{T}^{(\beta)}_{lm}}
{\mathbb{C}^{(\beta)}_{l}} \label{cutTT}
\end{eqnarray}
following Eq. (\ref{TT}), and a harmonic mode of the Gaussian bias
is estimated as
\begin{eqnarray}
\mathbb{N}^{TT(\alpha\times\beta)}_{LM} & = &
\frac{A^{TT(\alpha\times\beta)}_L}{\sqrt{L(L+1)}}
\displaystyle\sum_{ll'mm'}(-1)^Mf^{TT}_{lLl'} \left(
\begin{array}{ccc}
l'   &  l &   L\\
m' & m& -M
\end{array}
\right)
\nonumber \\
& \times &
\frac{\tilde{\mathbb{T}}^{(\alpha)}_{l'm'}}{\mathbb{C}^{(\alpha)}_{l'}}
\frac{\tilde{\mathbb{T}}^{(\beta)}_{lm}}{\mathbb{C}^{(\beta)}_l}
\label{cutNN}.
\end{eqnarray}
Here $\mathbb{C}$ are the power spectra of the observed lensed
temperature, determined from
$\langle\mathbb{T}_{lm}\mathbb{T}_{l'm'}\rangle$. As was done in
\cite{ACTlensing} and \cite{reion} we use the same power spectra in
Eq. (\ref{cutTT}) and Eq. (\ref{cutNN}). In order to deal with the
non-uniform noise distribution in the WMAP data, we symmetrize
$\mathbbm{d}^{TT(\alpha\times\beta)}_{LM}$ as in \cite{hirata2008},
denoting the symmetrized cross-correlation ``$TT(\alpha \bullet
\beta)$" between these two temperature maps,\begin{eqnarray}
\mathbbm{d}^{TT(\alpha\bullet\beta)}_{LM}&=&\frac{\mathbbm{d}^{TT(\alpha\times\beta)}_{LM}
+\mathbbm{d}^{TT(\beta\times\alpha)}_{LM}}{2}\label{rawsignal}
\end{eqnarray} and
\begin{eqnarray}
\mathbbm{N}^{TT(\alpha\bullet\beta)}_{LM}=\frac{\mathbbm{N}^{TT(\alpha\times\beta)}_{LM}
+\mathbbm{N}^{TT(\beta\times\alpha)}_{LM}}{2}.\label{gaussianbias}
\end{eqnarray}

We refer to $C_L^{est}=\langle d_{LM}^{\ast}d_{LM}\rangle$ as the
reconstruction including noise, and $N_L^{(0)}=\langle
N_{LM}^{\ast}N_{LM}\rangle$ as the Gaussian bias, with the
superscript ``$TT(\alpha \bullet \beta)$" omitted. Thus we obtain
\begin{widetext}
\begin{eqnarray}
\mathbbm{d}^{TT(\alpha\bullet\beta)}_{LM}&=&\frac{1}{2}\Bigg\{\frac{A^{TT(\alpha\times\beta)}_L}
{\sqrt{L(L+1)}}[\beta_{L0}\int d\mathbf{n} {\ }_{+1}Y^{\ast}_{LM}{\
}_0A^{T(\alpha)}X^{(\beta)}+\alpha_{L0}\int d\mathbf{n} {\
}_{-1}Y^{\ast}_{LM}{\ }_0A^{T(\alpha)}Y^{(\beta)}]\nonumber\\
&+&\frac{A^{TT(\beta\times\alpha)}_L}{\sqrt{L(L+1)}}[\beta_{L0}\int
d\mathbf{n} {\ }_{+1}Y^{\ast}_{LM}{\
}_0A^{T(\beta)}X^{(\alpha)}+\alpha_{L0}\int d\mathbf{n} {\
}_{-1}Y^{\ast}_{LM}{\ }_0A^{T(\beta)}Y^{(\alpha)}]\Bigg\},
\label{fastform}
\end{eqnarray}

\begin{equation}
\mathbbm{A}_L^{TT(\alpha\times\beta)} = \int^{-1}_{+1} d(\cos\theta)
\bigl[ \bigl(
\xi^{T(\alpha)}_{00}(\theta)\xi^{T(\beta)}_{11}(\theta)
-\xi^{T(\alpha)}_{01}(\theta)\xi^{T(\beta)}_{01}(\theta) \bigr)
d^L_{-1-1}(\theta) + \bigl(
\xi^{T(\alpha)}_{00}(\theta)\xi^{T(\beta)}_{1-1}(\theta)
-\xi^{T(\alpha)}_{01}(\theta)\xi^{T(\beta)}_{0-1}(\theta) \bigr)
d^L_{1-1}(\theta) \bigr],\label{fastformnoise}
\end{equation}
\end{widetext}
following a reasoning similar to the one near the end of Section II.
\begin{figure}
\centering
\includegraphics[width=9cm,height=9cm]{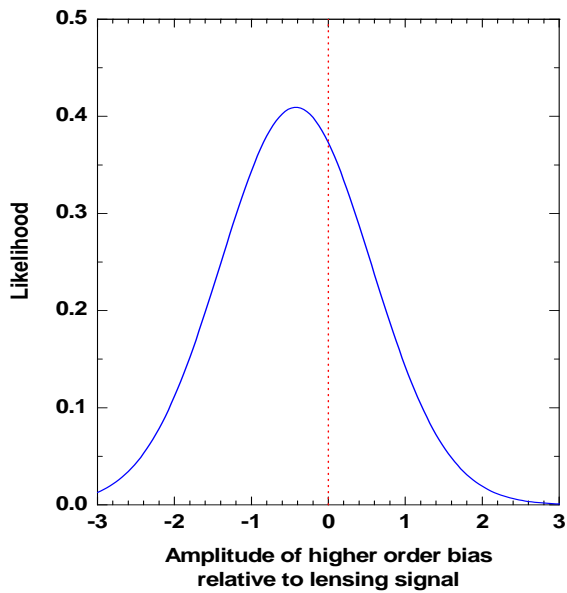}
\caption{The normalized likehood of the amplitude of the higher
order bias limited to the region $20<L<170$, to the simulated
lensing signal. This confirms that the higher order bias is
consistent with zero and negligible.} \label{higherlike}
\end{figure}

The two-point correlation of the Gaussian bias estimator is
essentially a four-point correlation function of the primordial
temperature modes. It should be carefully subtracted since, for a
noise-dominated experiment such as WMAP, the Gaussian four-point
bias is several orders of magnitude larger than the lensing power
spectra. In \cite{ACTlensing} phase-randomized data maps are used to
simulate this Gaussian bias. However, this approach does not work
for the present lensing reconstruction since WMAP's noise is not
isotropic. Evidence for this can be seen from the normalization
factor $\mathbbm{A}_L^{TT(\alpha\times\beta)}$ which is not equal to
$N^{(0)TT(\alpha\bullet\beta)}_{L}$ whereas they should be equal for
isotropic noise \cite{hufull}. The normalization factor Eq.
(\ref{fastformnoise}) only contains the partial contribution coming
from the non-isotropic noise while the Gaussian bias squared from
Eq. (\ref{gaussianbias}) consists of all the correlations generated
by the non-isotropic noise, see \cite{cooray} and \cite{KCK}. If the
phases of the WMAP temperature maps are randomized in order to
remove the lensing-induced coupling between modes, it will also
remove the strong correlation of the noise. The Gaussian bias
calculated in this way will be significantly lower than that from
the standard approach \cite{joeNG}. So we have to perform
simulations which use the simulated WMAP noise and temperature maps,
rather than the randomized WMAP data to get the Gaussian bias term.

The deflection power spectrum is
\begin{equation}
C_L^{dd} = \bigl\langle \bigl[ \mathbbm{d}^{TT(\alpha \bullet
\beta)}_{LM} \bigr]^\ast \, \mathbbm{d}^{TT(\alpha \bullet
\beta)}_{LM} - \bigl[ \mathbbm{N}^{TT(\alpha \bullet \beta)}_{LM}
\bigr]^\ast \, \mathbbm{N}^{TT(\alpha \bullet \beta)}_{LM}
\bigr\rangle. \label{deflectPS}
\end{equation}

\begin{figure}
\centering
\includegraphics[width=9cm,height=9cm]{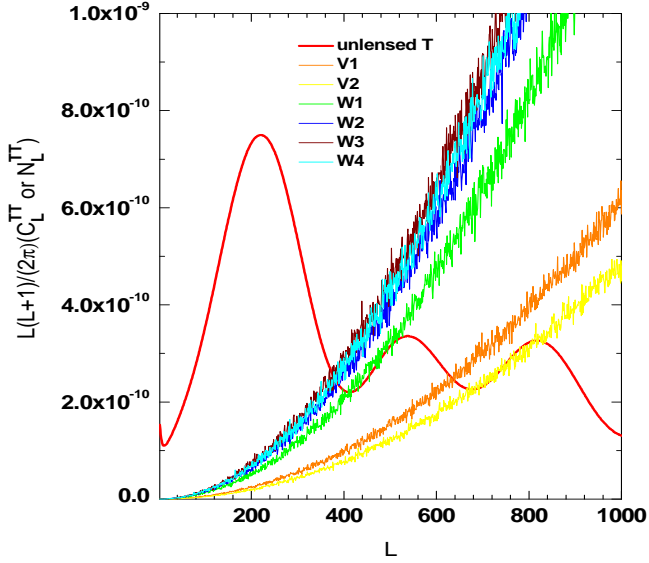}
\caption{WMAP noise for each DA and the $TT$ power spectrum as a
function of $L$.} \label{ttnoise}
\end{figure}
\begin{figure}
\centering
\includegraphics[width=9cm,height=9cm]{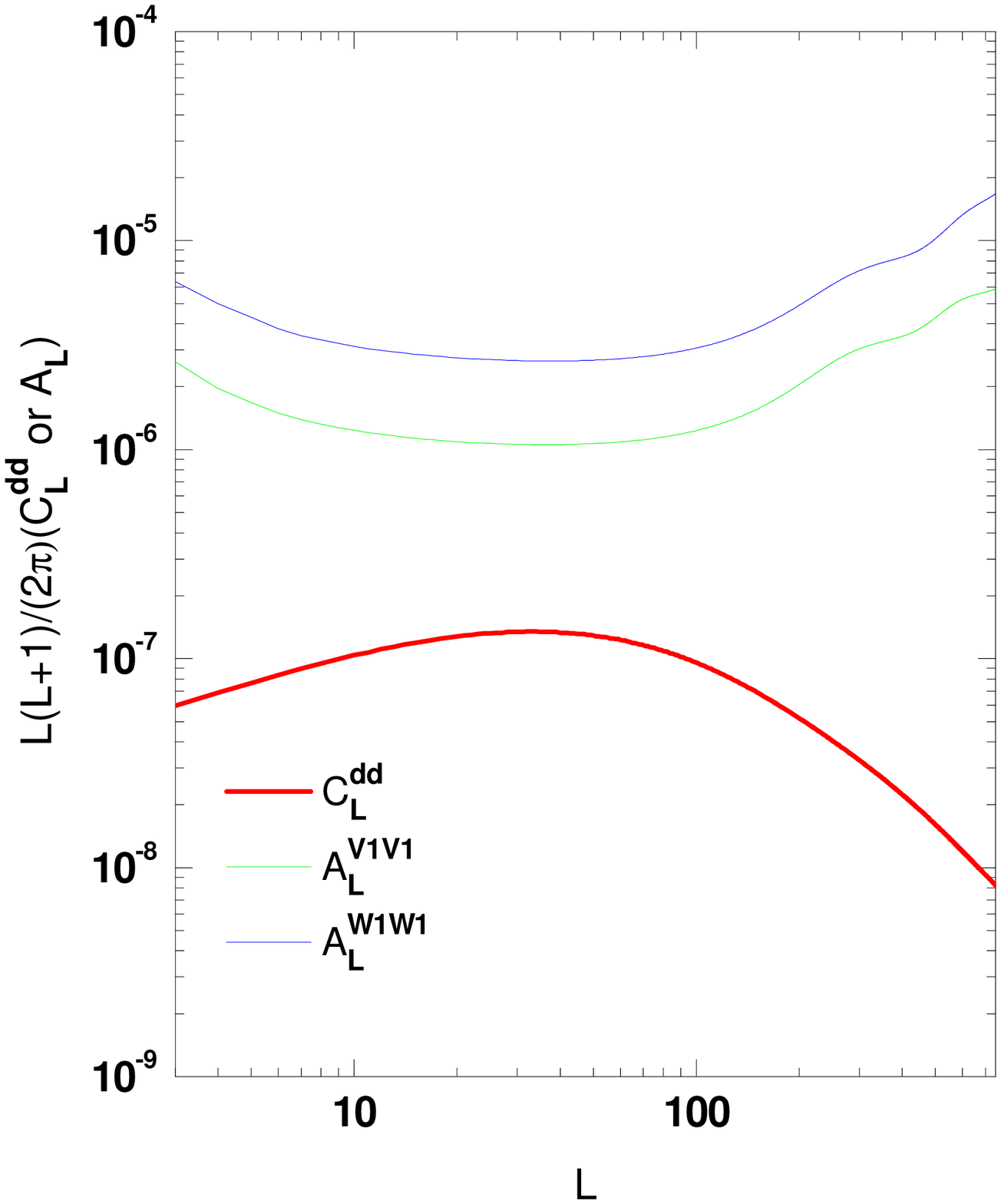}
\caption{Comparison of $\mathbb{A}_L$ (Eq. (\ref{fastformnoise}))
and the expected lensing signal as function of $L$. The estimator
noise is about two orders of magnitude higher that the signal
$C_L^{dd}$, indicating the difficulty of detecting lensing from
WMAP-7 data alone. } \label{estnoise}
\end{figure}
\begin{figure}
\centering
\includegraphics[width=9cm,height=9cm]{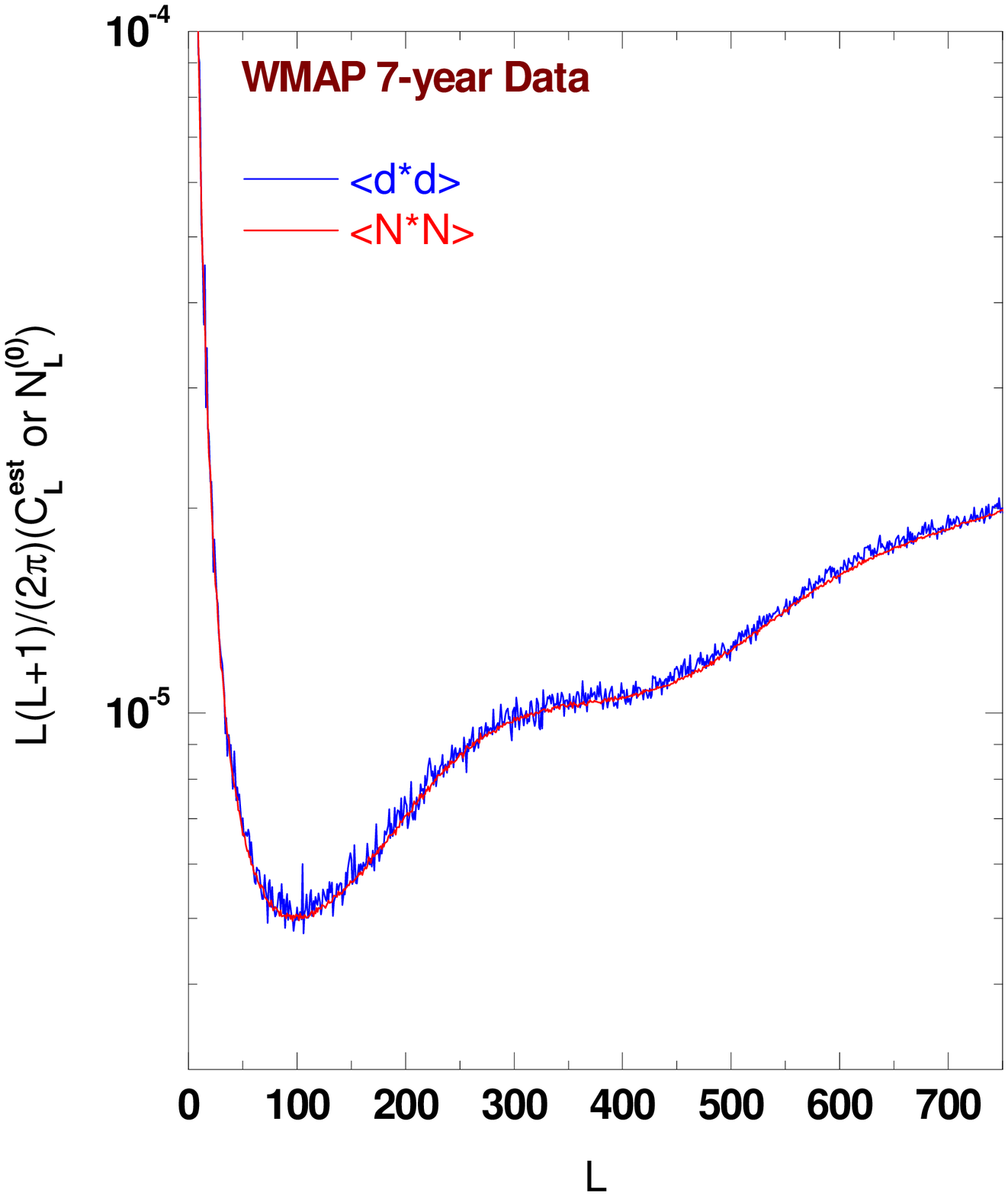}
\caption{The averaged reconstruction including noise ($C_L^{est}$)
(blue) of \textit{WMAP data} and the Gaussian bias $N^{(0)}_L$ (red)
from 700 realizations. Since lensing is approximately 100 times
smaller than $C_L^{est}$, the two curves are almost
indistinguishable; however, this confirms the precision of the noise
model. } \label{VWnoise}
\end{figure}
\begin{figure}
\centering
\includegraphics[width=9cm,height=9cm]{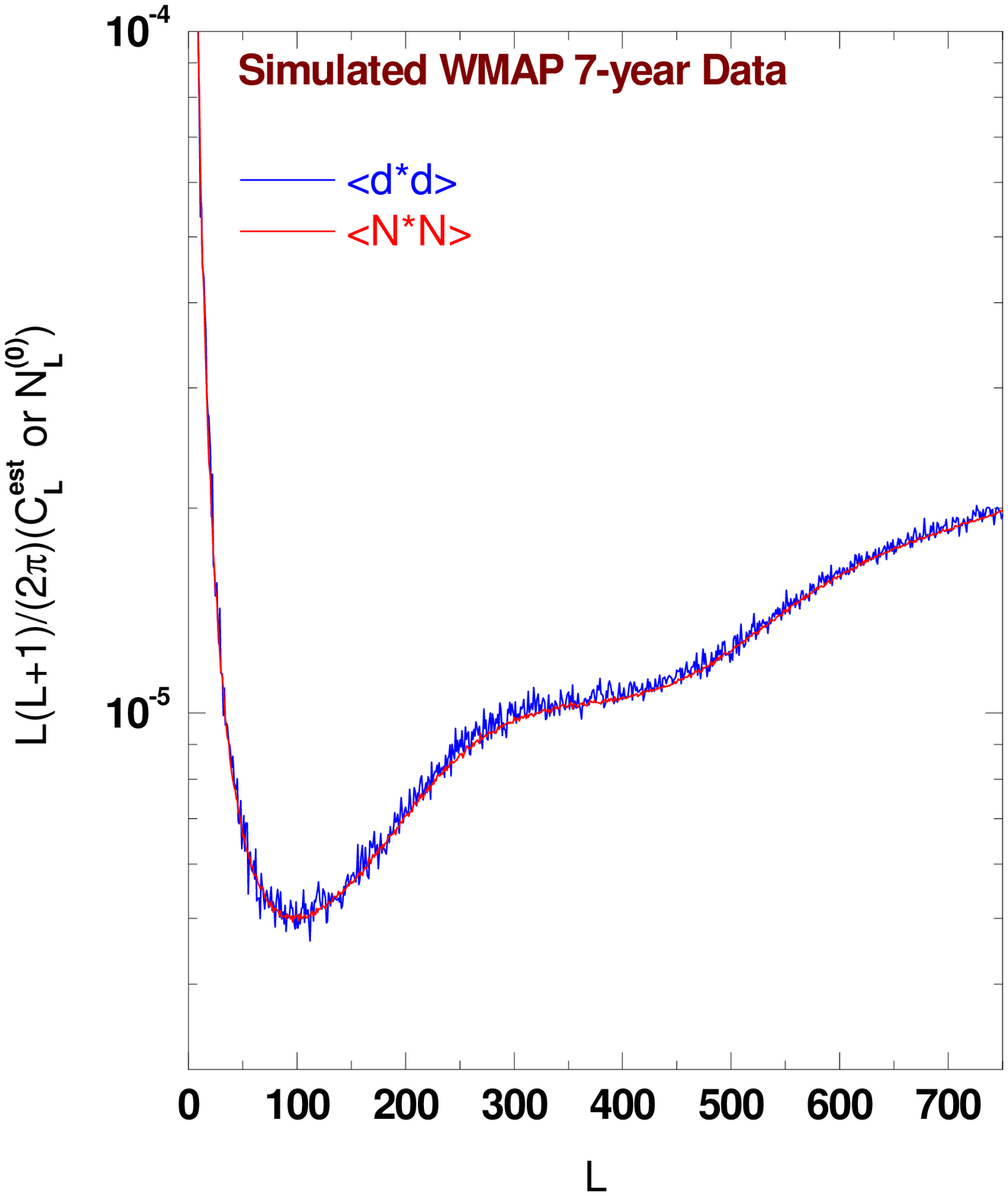}
\caption{The averaged reconstruction including noise ($C_L^{est}$)
(blue) of \textit{simulated WMAP data} and the Gaussian bias
$N^{(0)}_L$ (red) from 700 realizations. Since lensing is
approximately 100 times smaller than $C_L^{est}$, the two curves are
almost indistinguishable; however, this confirms the precision of
the noise model. } \label{SimVWnoise}
\end{figure}
\begin{figure}
\centering
\includegraphics[width=9cm,height=9cm]{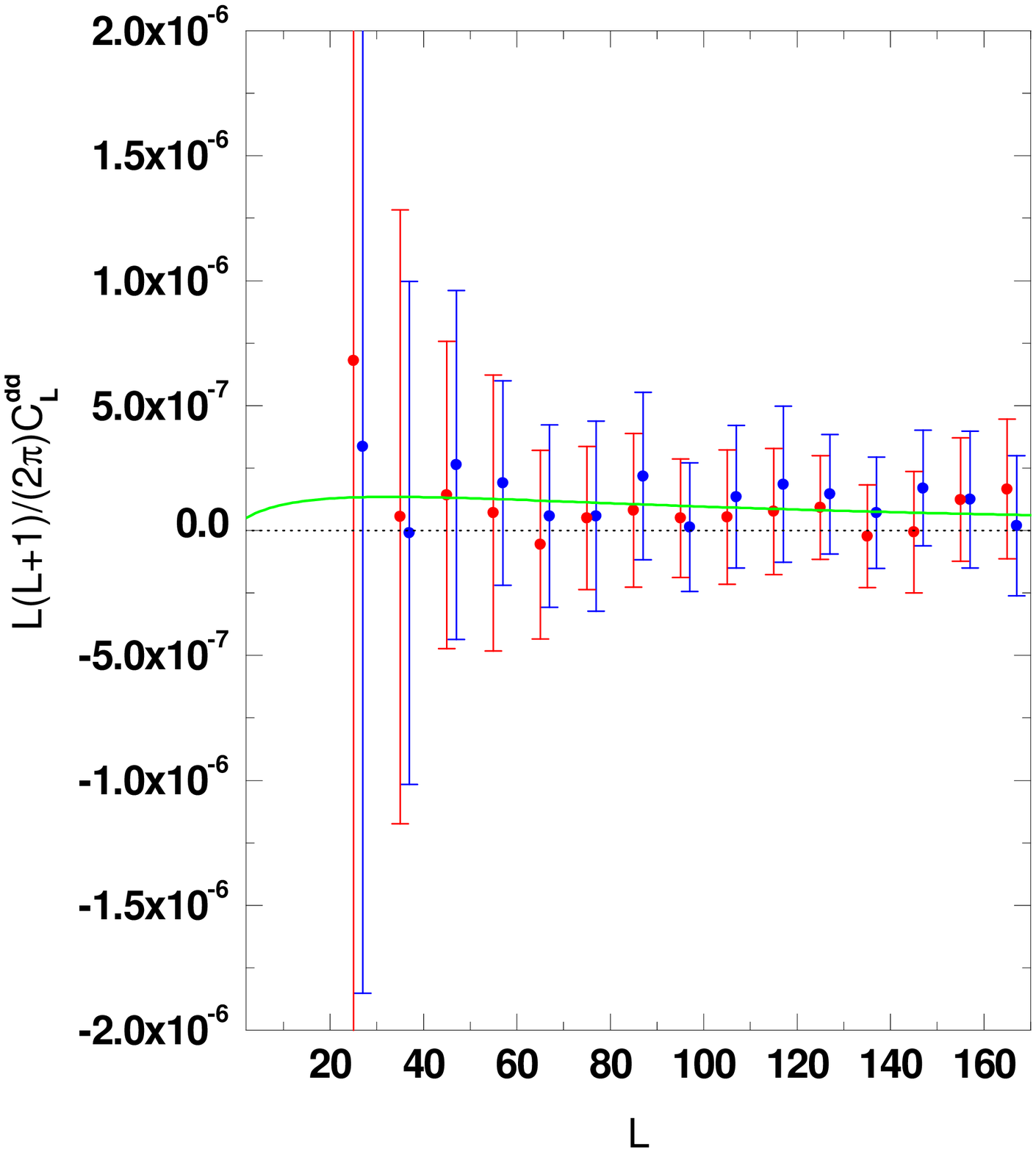}
\caption{The reconstructed power spectra ($C_L^{dd}$) of the
deflection angle field from all correlations of WMAP's W- and V-band
DAs. The green curve is the simulated lensing signal, and the data
points are the reconstructed lensing signal from simulations (red),
and the reconstructed lensing signal from data (blue). The red and
blue data points show the consistency between the simulated and real
WMAP data for the lensing reconstruction. } \label{VW}
\end{figure}
\begin{figure}
\centering
\includegraphics[width=9cm,height=9cm]{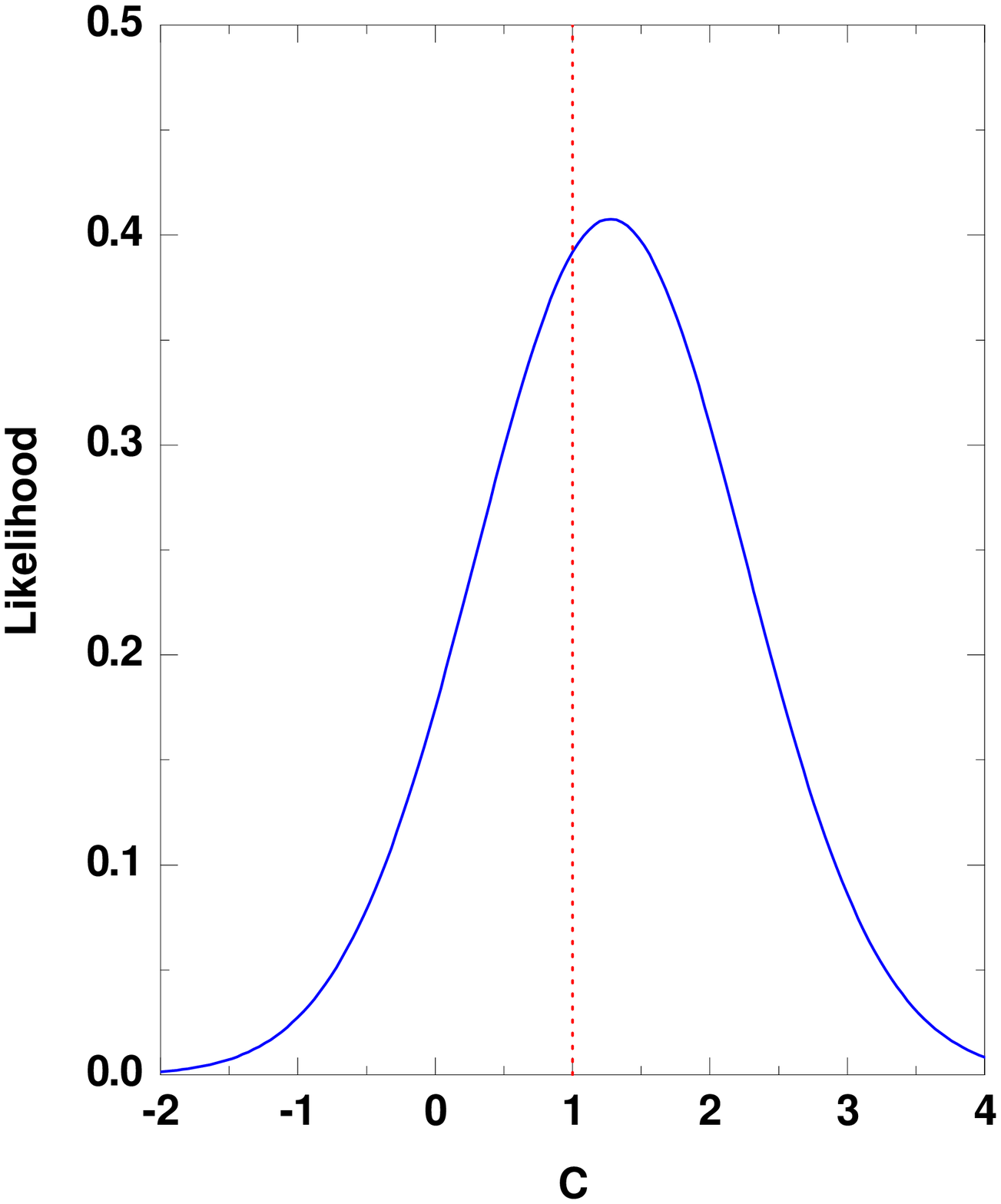}
\caption{The normalized likelihood distribution for $\mathcal{C}$
for all 21 correlations of WMAP's W- and V-band DAs.} \label{like}
\end{figure}
\begin{figure}
\centering
\includegraphics[width=9cm,height=9cm]{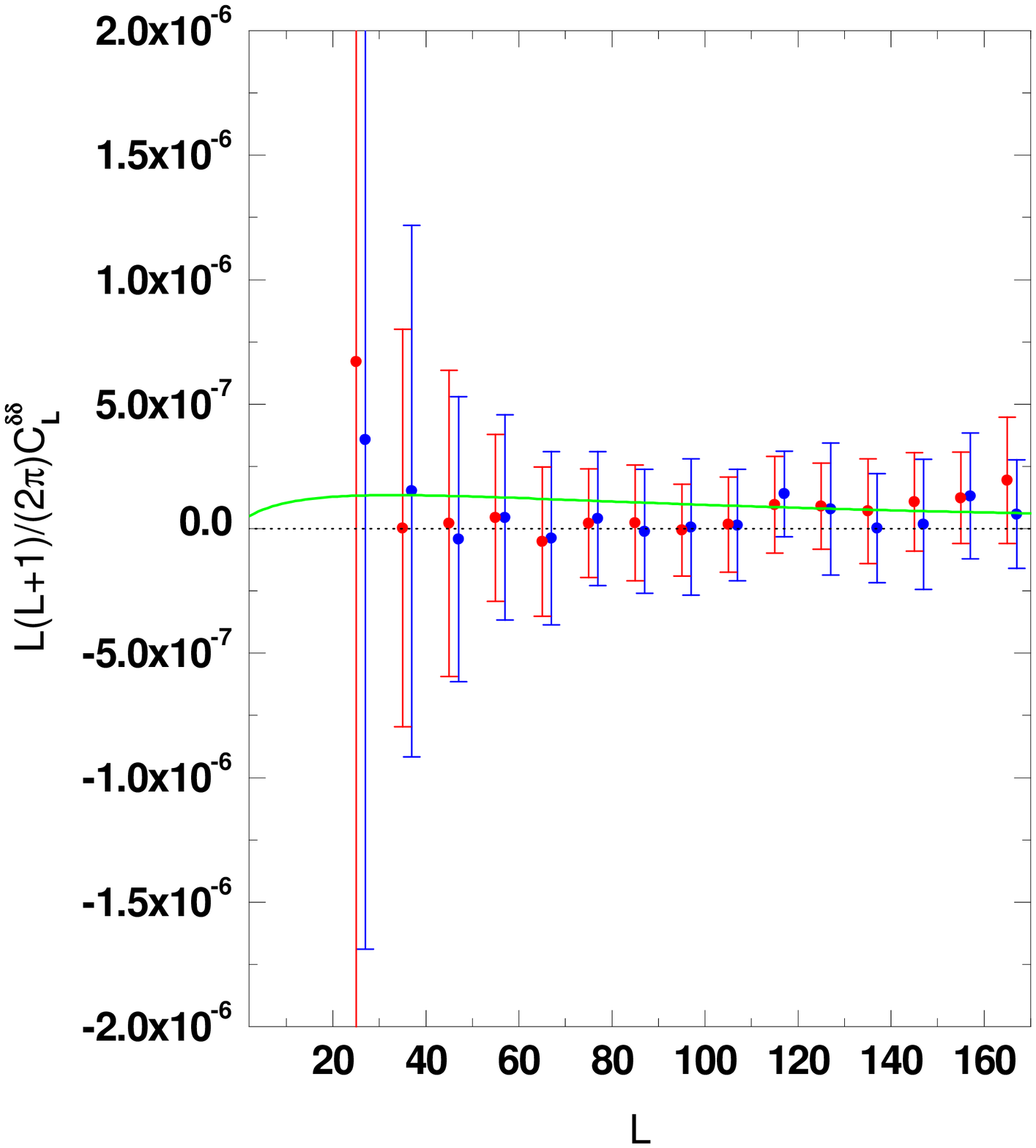}
\caption{Curl null test for all correlations of WMAP's W- and V-band
DAs: $C_L^{\delta\delta}$ from the simulated WMAP data (red), and
$C_L^{\delta\delta}$ from the real WMAP data (blue), for comparison,
the simulated lensing signal $C_L^{dd}$ (solid green). The red and
blue data points show the consistency between the simulated and the
real WMAP data for the curl null test. } \label{VWnull}
\end{figure}

This estimator is essentially the same as in \cite{ACTlensing}
except that here it is the full-sky version and the noise
$\mathbb{N}_{LM}$ is not obtained from the phase-randomized data. We
subtract the Gaussian bias for each realization of the estimator,
and all the estimated power spectra are averaged to get the binned
power spectra $\langle C_b^{dd} \rangle$ for the $b$-th bin
\cite{reion}. The averaged power spectrum in a range of $L$ labeled
by the index $b$ is
\begin{equation}
C_b^{dd} = \displaystyle\sum_{L\in b} \frac{L(L+1)}{b(b+1)}
C_L^{dd}.
\end{equation}The statistical uncertainty is given by $\sigma_b = [ \langle (C_b -
\bar C_b)^2 \rangle ]^{\frac{1}{2}}$. After the subtraction of the
Gaussian bias, there remains the higher order biases% which needs to be removed as well \oz{You make the case later that it does NOT need to be removed
, see \cite{ACTlensing} (where it was called ``null bias"),
\cite{reion}, and \cite{Mwhite}.

We expand $T$ in harmonic space as
\begin{eqnarray}
T_{LM} = \tilde T_{LM} + \delta T_{LM} + \delta^2 T_{LM} + \delta^3
T_{LM} + ..., \label{Texpansion}
\end{eqnarray}
see \cite{duncan}. Here the power $n$ in $\delta^n$ denotes the
order in $\phi^n$. We expand the noise bias as
\begin{eqnarray}
N_{L} = N_L^{(0)} + N_L^{(1)} + N_L^{(2)} + ..., \label{Nexpansion}
\end{eqnarray}
where the index $n$ in $N^{(n)}$ denotes the order of its dependence
upon $[\phi^{2}]^n$, excluding terms that contribute to the lensed
power spectrum. The four-point function $\langle
\mathbbm{d}^{\ast}_{LM} \mathbbm{d}_{LM}\rangle$ contains terms of
different order in $\delta^n T$. A term of the type $\langle \delta
T\delta T \tilde T \tilde T \rangle$ contributes to $C_L^{dd}$ and
the first order noise $N^{(1)}_L$ while terms of the type $\langle
\delta T \delta T \delta T \delta T \rangle$, $\langle \delta^2 T
\delta^2 T \tilde T \tilde T \rangle$, $\langle \delta^2 T \delta T
\delta T \tilde T \rangle$, and $\langle \delta^3 T \delta T \tilde
T \tilde T \rangle$ generate the second order noise $N_L^{(2)}$.
Following \cite{duncan}, the higher order bias term is calculated as
the difference between the estimated power spectrum and the sum of
its prediction and the lowest order noise (i.e., Gaussian bias):
$C_{L}^{est} - (C_L^{dd} + N^{(0)}_L)$, using Monte Carlo
simulations.

We study the statistical significance of the detection as follows.
Following \cite{hirata2008}, the reconstructed power spectra
$C^{(\textrm{obs})}$ are compared with their theoretical prior
$C^{(\textrm{th})}$ by minimizing a $\chi^2$ defined as
\begin{eqnarray}
\chi^2(\mathcal {C}) = \sum_{AB} (C^{(\textrm{obs})}_A - \mathcal{C}
C^{(\textrm{th})}_A) {\bf C}^{-1}_{AB} (C^{(\textrm{obs})}_B -
\mathcal{C} C^{(\textrm{th})}_B ) \label{chisq}
\end{eqnarray}
and varying $\mathcal{C}$. Here $A$ or $B$ label the range in $L$,
and $C_A$ or $C_B$ is the band-power. The covariance matrix
$\mathbf{C}$ is calculated from the Monte Carlo simulation as
$\mathbf{C}_{AB} = \langle(C^{(\textrm{sim})}_A -
\bar{C}^{(\textrm{sim})}_A) (C^{(\textrm{sim})}_{B} -
\bar{C}^{(\textrm{sim})}_{B})\rangle$. The best fit $\mathcal{C}$ is
obtained by setting the derivative of $\chi^2$ to zero:
\begin{equation}
\mathcal {C} = \frac{\sum_{AB} C^{(\textrm{th})}_A
\mathbf{C}^{-1}_{AB} C^{(\textrm{obs})}_{B}}
{\sum_{AB}C^{(\textrm{th})}_A\mathbf{C}^{-1}_{AB}C^{(\textrm{th})}_{B}}.\label{CL}
\end{equation}
A non-zero value of $\mathcal{C}$ indicates the presence of lensing.
The signficance of a non-zero value can be judged if its variance is
known. The variance of $\mathcal{C}$ is given by
\begin{equation}
(\Delta\mathcal {C})^2 = \frac{1}{\sum_{AB}C^{(\textrm{th})}_A
\mathbf{C}^{-1}_{AB} C^{(\textrm{th})}_{B}} \label{CLerror}
\end{equation}
and the significance of the detection of lensing is $\mathcal
{C}/\Delta \mathcal {C}$.

We show the higher order bias in Figure \ref{higher}. The higher
order bias $N_L^{(1)} + N_L^{(2)} + ...$ is seen to be negative for
$L < 20$ and positive for $L > 170$ and consistent with zero for $20
< L < 170$ where the amplitude is $-0.42\pm 0.98$ $(0.43\sigma)$,
compared to the simulated lensing signal $C_L^{dd}$ by using 15 bins
with $\Delta L = 10$ starting from $ L = 20$. In Figure
\ref{higherlike}, the likelihood of the amplitude of the higher
order bias limited to the region $20<L<170$ confirms that the bias
is consistent with zero. Thus subtraction of the higher order bias
is not required as long as we limit $L$ to this region.

\section{WMAP 7-year Data}

The lensing reconstruction depends most sensitively on the high-$L$
modes which are supplied by WMAP's DAs in the V (2 DAs) and W (4
DAs) frequency bands. Thus we use WMAP's coadded temperature maps
with r9 resolution (Healpix's $n_{side} = 512$) using all possible
distinct pairings: three auto-correlations for the two V-band DAs,
ten auto-correlations for the four W-band DAs, and eight
cross-correlations between the W- and V-band DAs for a total of 21
correlations (labeled ``ALL"). Smith et al. \cite{smith2007}, used
the Q-band DAs in addition to the W- and V-band DAs of WMAP 3-year
temperature maps. Hirata et al. \cite{hirata2008}, used 153 one-year
DAs from the WMAP 3-year data in the W- and V-bands. Recently, Smidt
et al. \cite{smidt2010} used the W- and V-frequency bands of the
WMAP 7-year data. This work adopts six DAs of WMAP's 7-year
temperature map, making the data selection slightly different from
other work, although the same signal-to-noise is expected. The WMAP
temperature maps contain very high levels of noise as shown in
Figure \ref{ttnoise}. The normalization factor $\mathbb{A}_L$ shown
in Figure \ref{estnoise} is about two orders of magnitude higher
than the signal $C_L^{dd}$; indicative of the difficulty of
extracting the lensing from the noisy data. We calculate the noise
in each band from WMAP's data instead of using an analytical form as
\cite{Hirata2004} and \cite{hirata2008} do. The noise is simulated
according to the prescription in \cite{joeNG}, and the beam transfer
functions are supplied by WMAP.

\begin{table}
\caption{Measurements of lensing $\mathcal{C}$ and its significance
$\mathcal{C} / \Delta \mathcal{C}$.}
\begin{tabular}{c|c|c}
\hline\hline
Data set & $\mathcal {C}$ & \hskip 2mm $\mathcal{C} / \Delta \mathcal{C}$ \\
\hline
WMAP-7 ALL\footnotemark[1] & $1.27\pm 0.98$ & $1.30\sigma$\\
\hline
WMAP-7 V$+$W\footnotemark[2] & $0.97\pm 0.47$ & $2.06\sigma$ \\
\hline
WMAP-1 ALL$\times$LRGs\footnotemark[3] & $1.0\pm 1.1$ & $0.91\sigma$ \\
\hline
WMAP-3 ALL$\times$(LRGs$+$QSOs$+$NVSS)\footnotemark[4] & $1.06\pm 0.42$ & $2.52\sigma$ \\
\hline
WMAP-3 (Q$+$V$+$W)$\times$NVSS\footnotemark[5] & $1.15\pm 0.34$ & $3.38\sigma$ \\
\hline
ACT\footnotemark[6] & $1.16\pm 0.29$ & $4.00\sigma$ \\
\hline
SPT\footnotemark[7] & - & $\sim4.90\sigma$ \\
\hline
\end{tabular} \\
\label{TCL} \footnotetext[1]{All 21 correlations of WMAP-7's W- and
V-band DAs in this work.} \footnotetext[2]{\cite{smidt2010} WMAP-7 V
and W bands. } \footnotetext[3]{\cite{Hirata2004} WMAP-1 W- and
V-band DAs, LRGs. } \footnotetext[4]{\cite{hirata2008} WMAP-3 W- and
V-band DAs, LRGs, QSOs and NVSS. } \footnotetext[5]{\cite{smith2007}
WMAP-3 Q-, V-, W-band DAs, NVSS. }
\footnotetext[6]{\cite{ACTlensing} ACT temperature maps. }
\footnotetext[7]{\cite{SPT} SPT temperature maps. }
\end{table}

\begin{table}
\caption{Summary of $\mathcal{C}$ and its significance $\mathcal{C}
/ \Delta \mathcal{C}$ for this work.}
\begin{tabular}{c|c|c}
\hline\hline
Type & $\mathcal {C}$ & \hskip 2mm $\mathcal{C} / \Delta \mathcal{C}$ \\
\hline
higher order bias& $-0.42\pm 0.98$ & $0.43\sigma$\\
\hline
curl null test& $0.38\pm 0.79$ & $0.47\sigma$ \\
\hline reconstructed lensing& $1.27\pm 0.98$ & $1.30\sigma$\\ \hline
\end{tabular} \\
\label{summary}
\end{table}

\section{Simulation and Analysis}

We use the CAMB code \cite{camb} to obtain the power spectra
$\tilde{C}_l^{TT}$, and $C_l^{\phi\phi}$ using a six parameter
$\Lambda$CDM model with $\textbf{P} = {\omega_bh^2, \omega_ch^2, h,
\tau, A_s, n_s} = {0.0226, 0.112, 0.70, 0.09, 2.1 \times 10^{-9},
0.96}$. These are input into a pipeline that has elements as
follows.

Gaussian maps of the deflection angle field $\mathbf{d}(\mathbf{n})$
and unlensed temperature $\tilde{T}(\mathbf{n})$ are generated using
their respective power spectra. We use $\tilde{C}_l^{TT}$, and
$C_l^{\phi\phi}$ to create one realization of the simulated
deflection field and lensed temperature maps $T(\mathbf{n})$ are
generated using Eq. (\ref{cmblensing}) with the
$\tilde{T}(\mathbf{n})$ and $\mathbf{d}(\mathbf{n})$ found above.

Using the inverse SHT, the temperature maps are converted into
harmonic modes $a_{lm}$ which are convolved with the beam transfer
functions $b_l$. Using SHT, these are transformed into configuration
space and WMAP-based noise is added. We mask the galactic plane and
point sources using WMAP's KQ75 mask. Using inverse SHT, the
resulting maps are transformed back into harmonic space where new
$a_{lm}$ are kept up to $l_{max} = 750$ and $|m_{max}|=750$.

The noise simulation is crucial to this work because the $N^{(0)}_L$
is one hundred times larger than $C_L^{dd}$. The six V- and W-band
DAs, labeled by $\alpha = V1$, $V2$, $W1$, $W2$, $W3$, $W4$, have
different noise variances, different beam transfer functions, and
different relative phases. To mimic the WMAP DAs, we simulate the
Gaussian bias as follows. Using
\begin{eqnarray}
\tilde{\mathbb{T}}^{(i)\alpha}(\mathbf{n}) &=& M(\mathbf{n}) \Bigl[
\int d\mathbf{n}' \, \tilde{T}^{(i)}(\mathbf{n}')
B^{\alpha}(\mathbf{n}, \mathbf{n}') \nonumber\\&+&
N^{(i)\alpha}(\mathbf{n}) \Bigr],\label{SimGaussian}
\end{eqnarray}
we set the index $i$ (an arbitrary running index) for Eq.
(\ref{SimGaussian}) and generate an unlensed temperature map
$\tilde{T}^{(i)}(\mathbf{n})$, and six noise maps
$N^{(i)\alpha}({\bf n})$, $\alpha = V1$, $V2$, $W1$, $W2$, $W3$,
$W4$. Then we make an observed map
$\tilde{\mathbb{T}}^{(i)\alpha}(\mathbf{n})$ using Eq.
(\ref{SimGaussian}), and repeat this procedure to make another
observed map $\tilde{\mathbb{T}}^{(i)\beta}(\mathbf{n})$.
Subsequently, we calculate the Gaussian bias $N^{(0)}_L$ using Eq.
(\ref{gaussianbias}) for the pair $(\alpha \bullet \beta)$. In the
same way, we generate 21 realizations for all the correlations.
Finally we increase the index $i$, and repeat the whole procedure
until the ensemble $\{N^{(0)}_L\}$ has 700 elements.

We proceed in a similar manner simulating the reconstruction
including noise, except setting $T^{(i)}({\bf n})=T({\bf n})$ and
$N^{(i)\alpha}(\mathbf{n})=N^{\alpha}(\mathbf{n})$. Using
\begin{eqnarray}
\mathbb{T}^{(i)\alpha}(\mathbf{n}) &=& M(\mathbf{n}) \Bigl[ \int
d\mathbf{n}' \, T^{(i)}(\mathbf{n}') B^{\alpha}(\mathbf{n},
\mathbf{n}') \nonumber\\&+& N^{(i)\alpha}(\mathbf{n})
\Bigr],\label{SimSignal}
\end{eqnarray}
we set the index $i$ for Eq. (\ref{SimSignal}), and generate a
lensed temperature map $T^{(i)}(\mathbf{n})$, and six noise maps
$N^{(i)\alpha}({\bf n})$, $\alpha = V1$, $V2$, $W1$, $W2$, $W3$,
$W4$. Then we make an observed map
$\mathbb{T}^{(i)\alpha}(\mathbf{n})$ using Eq. (\ref{SimSignal}) and
repeat this procedure to make another observed map
$\mathbb{T}^{(i)\beta}(\mathbf{n})$. Subsequently we calculate the
reconstruction including noise $C^{est}_L$ using Eq.
(\ref{rawsignal}) for the pair $(\alpha \bullet \beta)$. In the same
way, we generate 21 realizations for all the correlations. Finally
we increase the index $i$, and repeat the whole procedure until the
ensemble $\{C^{est}_L\}$ has 700 elements. Eq. (\ref{deflectPS}) is
then used to obtain the deflection power spectrum $C_L^{dd}$.

We show the reconstruction including noise $C_L^{est}$ and the
Gaussian bias $N_L^{(0)}$ in Figures \ref{VWnoise}, and
\ref{SimVWnoise} for the real and the simulated WMAP data,
respectively. The simulation is consistent with the data, and we
confirm that the two terms in Eq. (\ref{deflectPS}) nearly have the
same magnitude, and the lensing induced difference is not visible
because the lensing signal $C_L^{dd}$ is one hundred times smaller
than the Gaussian bias $N_L^{(0)}$. We use Eq. (\ref{deflectPS}) to
calculate the reconstructed lensing power spectra in Figure
\ref{VW}.

The likelihood distribution of $\mathcal{C}$ is shown in Figure
\ref{like}, where it is seen lensing is detected at
only 1.30$\sigma$ confidence level.\\\\

\section{Curl Null Test}

To check for systematic effects, we employ the ``curl null test".
The deflection angle field can be written as the sum of a gradient
and a curl term \cite{hirataPolar}:
\begin{equation}
D_i(\mathbf{n})=d_i(\mathbf{n})+\epsilon_{ij}\nabla^j\delta(\mathbf{n}).
\label{2part}
\end{equation}
The first term leads to the Hu estimator \cite{waynehuTri,
waynehuTT}
\begin{equation}
d^{TT}_{LM} = \frac{A^{TT}_L}{\sqrt{L(L+1)}} \int d\mathbf{n}\,
Y^{\ast}_{LM} \nabla^i \bigl[ {}_0A^T(\mathbf{n})\nabla_i \,
_0B^T(\mathbf{n}) \bigr] \label{grad}
\end{equation}
whose efficient form is given in Eq. (\ref{fastform}), here $_0
A^T({\bf{n}})$ is given by Eq. (\ref{abbrivA}) and
\begin{eqnarray}
_0B^T({\bf{n}}) = \displaystyle \sum_{lm}
\frac{\tilde{C}^{TT}_l}{C^{tot}_l} T_{lm} \, _{0} Y_{lm}({\bf{n}}).
\end{eqnarray}

The estimator for the curl part in Eq. (\ref{2part}) is
\begin{equation}
\delta^{TT}_{LM} = \displaystyle \sum_{ij} \epsilon^{ij}
\frac{A^{TT}_L}{\sqrt{L(L+1)}} \int d\mathbf{n} \, Y^{\ast}_{LM}
\nabla_i \bigl[ _0A^T(\mathbf{n})\nabla_j \, _0B^T(\mathbf{n})
\bigr] \label{curl}
\end{equation}
and the corresponding efficient form is
\begin{widetext}
\begin{eqnarray}
\delta^{TT(\alpha\bullet\beta)}_{LM} & = & \frac{1}{2} \Bigg\{
\frac{A^{TT(\alpha\times\beta)}_L}{\sqrt{L(L+1)}} \biggl[
\beta_{L0}\int d\mathbf{n} {\ }_{+1}Y^{\ast}_{LM} \, _0A^{T(\alpha)}
X^{(\beta)} - \alpha_{L0} \int d\mathbf{n} \,
_{-1}Y^{\ast}_{LM}{\ }_0A^{T(\alpha)}Y^{(\beta)} \biggr] \nonumber \\
& + &\frac{A^{TT(\beta\times\alpha)}_L}{\sqrt{L(L+1)}} \biggl[
\beta_{L0}\int d\mathbf{n} \, _{+1}Y^{\ast}_{LM} \, _0A^{T(\beta)}
X^{(\alpha)}-\alpha_{L0}\int d\mathbf{n} \, _{-1}Y^{\ast}_{LM}{\
}_0A^{T(\beta)}Y^{(\alpha)} \biggr] \Bigg\}, \label{fastformCurl}
\end{eqnarray}
\end{widetext}
which can be compared with Eq. (\ref{fastform}). We show the
resulting power spectra $C_L^{\delta\delta}$, averaged from 700
realizations from the real and the simulated WMAP data separately in
Figure \ref{VWnull}. The averaged curl component amplitude is $0.38
\pm 0.79$ consistent with zero as expected, compared to the
simulated $C_L^{dd}$.

\begin{figure}
\centering
\includegraphics[width=9cm,height=9cm]{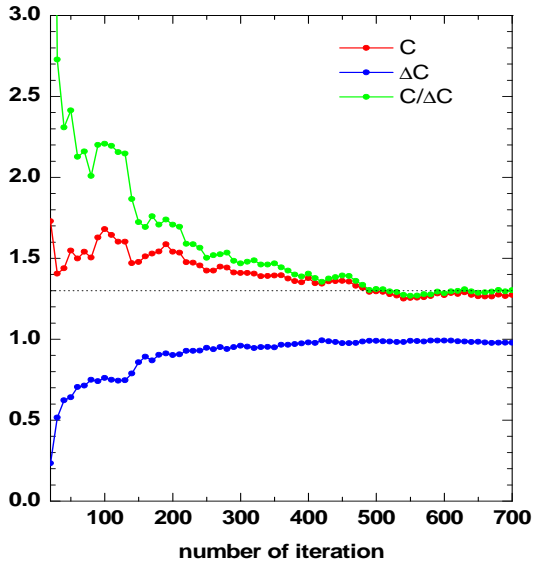}
\caption{The convergence behavior. The values of mean amplitude
$\mathcal {C}$ (red), the error $\Delta \mathcal {C}$ (blue), and
the detection significance $\mathcal {C}/\Delta \mathcal {C}$
(green) of the reconstructed lensing signal $C_L^{dd}$ are plotted
for every 10 realizations. It is seen that convergence is reached
after 700 realizations. } \label{converge}
\end{figure}

\section{Results and Discussion}

In this work, we have applied the optimal quadratic estimator to
WMAP-7 temperature maps alone for the first time.

We have monitored the convergence behavior for the mean value
$\mathcal {C}$, the error $\Delta \mathcal {C}$, and the detection
significance $\mathcal {C}/\Delta \mathcal {C}$ of the reconstructed
lensing signal $C_L^{dd}$. We find that all these quantities
converge after producing 700 realizations of the reconstructed
lensing signal, see Figure \ref{converge}. We determine the
significance of the lensing detection and find $\mathcal {C}=1.27\pm
0.98$ $(1.30\sigma)$, while Smidt et al. found $\mathcal {C}=0.97\pm
0.47$ $(2.06\sigma)$. The result is shown in Table \ref{TCL} as well
as a comparison with \cite{Hirata2004, hirata2008, smith2007,
ACTlensing, smidt2010, SPT}.  All our results have been corrected by
the sky fraction. We find evidence for lensing only at $1.30\sigma$,
using all correlations of WMAP-7's W- and V-band DAs. The resulting
constraint on the lensing amplitude differs from
 \cite{smidt2010} and this can be explicated from several aspects.
In terms of the estimator, we use the optimal estimator derived from
minimum variance principle \cite{waynehuTri}, rather than the
kurtosis estimator in \cite{smidt2010}. We adopt the individual beam
transfer function associated with each DA, not the averaged one for
each frequency. We have taken into account the impact of the higher
order bias, afterwards restricting the reconstruction in a proper
multiple range that marginally overlaps with \cite{smidt2010}. In
terms of the noise model, we estimate the noise in a way which
mimics WMAP, not simply generating random underlying skies and
associated noises with independent phases. All these factors may
jointly contribute to the difference between us and Smidt et al. A
summary of various tests in this work is shown in Table
\ref{summary}. We do not observe a significant lensing signal from
the WMAP 7-year temperature data.

We did not apply a correction for higher order bias terms
$N^{(1)}_L$, $N^{(2)}_L$, ..., because they are expected to be small
owing to the fact that we limited the region of $L$ to $20 < L <
170$, where the higher order bias is consistent with zero. The
higher order bias can be obtained via an iterative
solution\cite{KCK} but it is computationally demanding and not
warranted in the present case because we do not obtain a significant
signal.

We applied the curl null test to all the correlations of W- and
V-band DAs as a systematic check, since we observe a small amount of
power from the reconstructed gravitational lensing signal (Figure
\ref{VW}). The reconstruction procedure passes the curl null test.

The effects of beam systematics and the galactic and foreground
contaminations are quite small compared to the statistical error. We
do not correct the statistical result for the presence of point
sources because they introduce negligible systematics \cite{joeNG}.

We have demonstrated, using a nearly optimal estimator, that WMAP-7 data does not have the power to
detect gravitational lensing, which is unfortunate since WMAP data is
the only publicly available data set with sufficient angular
resolution to detect lensing. However, WMAP-7 does have value as a
publicly available tool to assess the efficacy of lensing algorithms
and to test for systematic biases.

\begin{acknowledgments}
We would like to acknowledge helpful discussions with Joseph Smidt,
Meir Shimon, Aneesh V. Manohar, Grigor Aslanyan, and Edward Wollack.
We acknowledge the use of CAMB, Healpix software packages.
\end{acknowledgments}

\end{document}